\documentclass[twocolumn]{revtex4}
\usepackage{graphicx}
\usepackage{dcolumn}
\usepackage{color}
\usepackage{bm}
\usepackage{amsmath}
\usepackage{amssymb}
\usepackage{graphicx,verbatim}
\usepackage{natbib}

\newcommand{\pd}[2]{\frac{\partial #1}{\partial #2}}
\newcommand{\mf}{\mathbf}

\newcommand{\bal}{\begin{align*}}
\newcommand{\eal}{\end{align*}}
\newcommand{\p}{\partial}

\newcommand{\pr}{\prime}

\newcommand{\bra}[1]{\left\langle #1 \right|}
\newcommand{\ket}[1]{\left|#1\right\rangle}

\renewcommand{\Im}{\operatorname{Im}}

\begin{document}

\title{Controlling the group velocity of colliding atomic Bose-Einstein condensates with Feshbach resonances}

\author{Ranchu Mathew}
\author{Eite Tiesinga}%
 \affiliation{Joint Quantum Institute, National Institute of Standards and Technology and University of Maryland, 
 College Park, 20740, USA.}

\date{\today}
            

\begin{abstract}
We report on a proposal to change the group velocity of a small Bose
Einstein Condensate (BEC) upon collision with another BEC in analogy
to slowing of light passing through dispersive media.  We make use of
ultracold collisions near a magnetic Feshbach resonance, which gives
rise to a sharp variation in scattering length with collision energy
and thereby changes the group velocity. A generalized Gross-Pitaveskii
equation is derived for a small BEC moving through a larger stationary
BEC. We denote the two condensates by laser and medium BEC, respectively,
to highlight the analogy to a laser pulse travelling through a medium. We
derive an expression for the group velocity in a homogeneous medium
as well as for the difference in distance, $\delta$, covered by the
laser BEC in the presence and absence of a finite-sized medium BEC with 
a Thomas-Fermi density distribution. For
a medium and laser of the same isotopic species, the shift $\delta$
has an upper bound of twice the Thomas-Fermi radius of the medium.
For typical narrow Feshbach resonances and a medium with number density
$10^{15}$ cm$^{-3}$ up to $85\%$ of the upper bound can be achieved,
making the effect experimentally observable. We also derive constraints
on the experimental realization of our proposal.

\end{abstract}

\maketitle

Over the last two decades significant advances have been made to
replicate linear and non-linear optical phenomena with matter
waves, creating the field of matter-wave optics. For example,
atom lasers \cite{mewes_output_1997, hagley_well-collimated_1999}
are sources of coherent ultra-cold atoms generated by extracting
atoms from a Bose-Einstein condensate (BEC).  The coherence
of BECs was demonstrated by interfering two condensates
\cite{andrews_observation_1997}. Atomic mirrors and beam splitters have
also been realized \cite{cronin_optics_2009}. Recently, the matter-wave
equivalent of meta-materials (media with negative refractive index)
has been proposed \cite{baudon_negative-index_2009}. The analog of
nonlinear four-wave mixing has been demonstrated using atom lasers
\cite{deng_four-wave_1999,kheruntsyan_violation_2012}. In these
experiments three BECs with phase-matched relative momenta generated a
fourth beam.

We present a proposal to slow a BEC while propagating 
through another BEC near a magnetic Feshbach resonance in
analogy to slowing of light in dispersive media. Slowing of light
occurs when the refractive index of a medium varies sharply with
photon frequency. Using electromagnetically induced transparency
\cite{fleischhauer_electromagnetically_2005}, slow light has been
observed with a Bose-Einstein condensate \cite{hau_light_1999,
ginsberg_coherent_2007} and a hot Rb gas \cite{kash_ultraslow_1999}
acting as the medium.

Magnetic Feshbach resonances are a tool
with which to manipulate the interaction between ultracold atoms
\cite{chin_feshbach_2010}. They are used for creating ultracold
molecules, molecular condensates , and in the BEC-BCS
crossover in Fermi gases \cite{inguscio_2008, giorgini_theory_2008}.  Feshbach resonances play an essential role
in condensing $^{133}$Cs, $^{85}$Rb and $^{39}$K
\cite{weber_bose-einstein_2003,cornish_stable_2000,roati_39k_2007}.
Cooling ultracold atoms using Feshbach resonances  has been
proposed \cite{mathey_collisional_2009}. Collisions can also be tuned
using optical Feshbach resonances \cite{ciurylo_optical_2005,
blatt_measurement_2011, yan_controlling_2009}, as their width can be
dynamically varied with a laser. They, however, tend
to suffer from losses due to spontaneous emission.

\begin{figure}
\includegraphics[width=.45\textwidth]{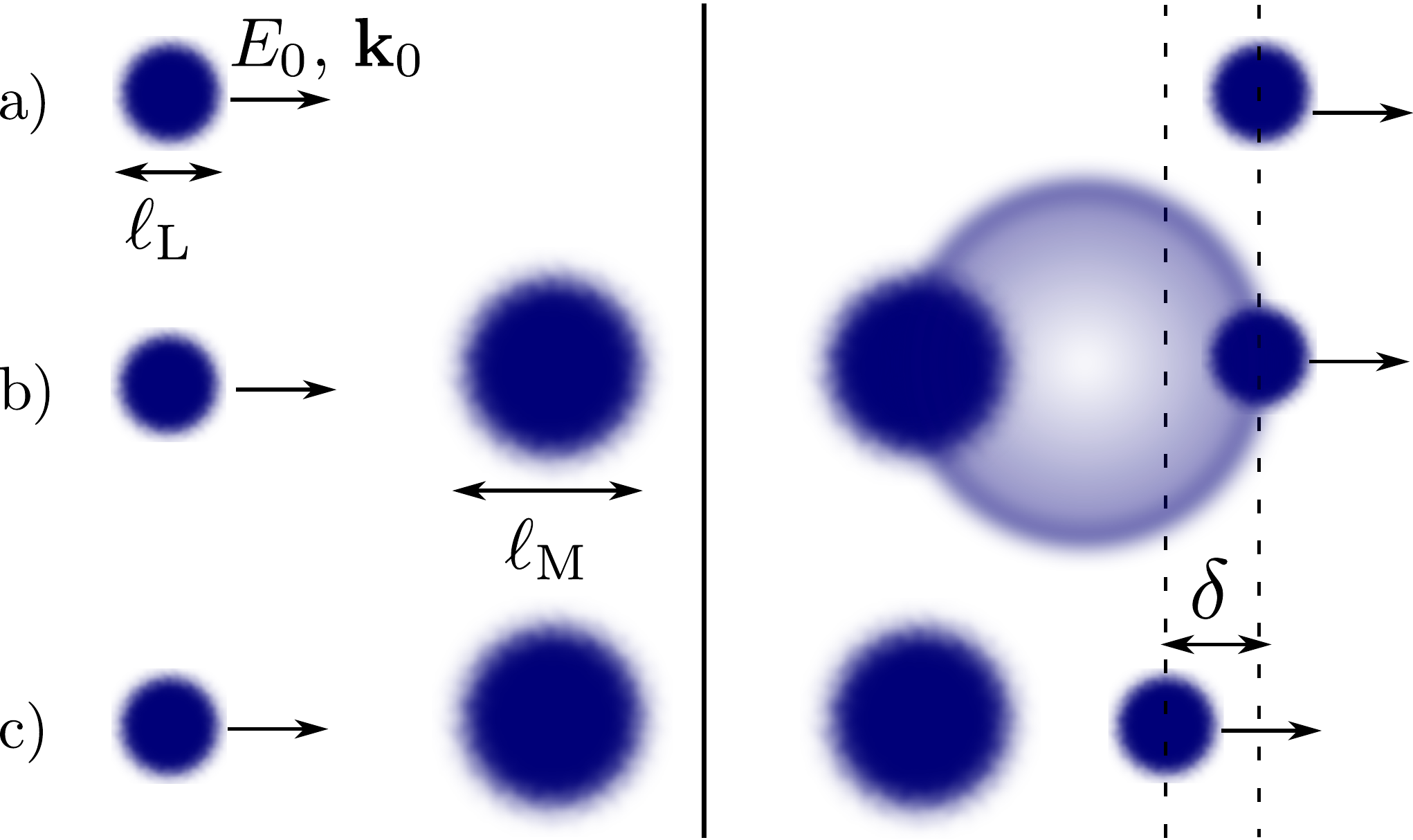}
\caption{A schematic of a slow-atom experiment with colliding
condensates. The left and right hand side show images of the condensates
before and after the collision, respectively. Panel a) simply shows the
laser BEC moving in free space with kinetic energy $E_0$ and 
wave vector $\mf{k}_0$. Panels b)
and c) show two cases of a laser BEC propagating through the medium BEC.
Panel b) shows a case with large elastic scattering losses indicated
by the halo of scattered atoms . Finally, panel c) depicts the
collision near a Feshbach resonance where there is negligible scattering
loss. This occurs when the scattering amplitude is zero. The distance
delay $\delta$ of the laser BEC is also indicated.  }
\label{fig_schematic} 
\end{figure}

Figure \ref{fig_schematic} shows our proposal of slowing a condensate
travelling through a large stationary BEC near a magnetic Feshbach
resonance. These two BECs play the role of a laser and a medium, respectively.
The laser BEC contains atoms of mass $m$ and has a diameter
$\ell_{\rm L}$. It propagates with wavevector $\mf{k}_0$ of magnitude $k_0
\gg 1/\ell_{\rm L}$. Its average velocity is $\mf{v}_0= \hbar \mf{k}_0/m$ and the
kinetic energy per particle is $E_0 = \hbar^2 k_0^2/(2m)$.  (Here $\hbar$
is the reduced Planck's constant.) It is incident on a stationary BEC
of size $\ell_{\rm M}$, such that $\ell_{\rm L}\ll \ell_{\rm M}$ (in all
spatial directions). Figures \ref{fig_schematic}b and c show two distinct
cases of collisions between condensates. In Fig.~\ref{fig_schematic}b
elastic scattering out of the two condensates is significant. 
These losses have limited four-wave mixing
experiments \cite{chikkatur_suppression_2000}. On the other
hand, it allows detection of $d$-wave shape resonances
in collisions between BECs \cite{buggle_interferometric_2004} and
thermal gases \cite{thomas_imaging_2004}.  In Fig.~\ref{fig_schematic}c
the collision takes place at the lossless point, where the
scattering cross section is small due to interference of the background
and resonance scattering amplitudes. As we will derive the laser
BEC then slows down. The slowing can be quantified by $\delta$, the
difference in distance travelled by the laser BEC in the presence and
absence of the medium.

We first review the two-body physics of a Feshbach
resonance. Subsequently, we derive the equation of motion for 
a laser BEC travelling through a medium.  We then derive the group
velocity in a homogeneous medium followed by an estimate
for $\delta$ for an inhomogeneous medium whose density is described
by a Thomas-Fermi profile.  Finally, restrictions on an experimental
realization are given.

Collisions between ultracold atoms are dominated by
$s$-wave scattering and the scattering amplitude $f(E)$ only depends on the relative
collision energy, $E = \hbar^2 k^2/(2\mu)$.
Then around an isolated magnetic Feshbach resonance
\cite{taylor_scattering_1983,kohler_production_2006,chin_feshbach_2010} 
\begin{equation}
f(E) = f_{\rm bg}(E)- e^{2 i \delta_{\rm bg}}
	\frac{\hbar\Gamma(E)/(2k)}{E  - E_{\rm res} +i\hbar\Gamma(E)/2}\,
	, \label{eq_sc_amp}
\end{equation} 
where $\mu$ is the reduced mass and $k$ is the magnitude
of relative wavevector. The background scattering amplitude $f_{\rm
bg}(E)= e^{i\delta_{\rm bg}}\sin(\delta_{\rm bg})/k$ with phase shift
$\delta_{\rm bg}$. To a good approximation $f_{\rm bg}(E)= -a_{\rm
bg}/(1+ika_{\rm bg})$, and $\delta_{\rm bg}= -k a_{\rm bg}$, where
$a_{\rm bg}$ is the background scattering length. The resonance width
$\hbar\Gamma(E)= 2 k a_{\rm bg}\Gamma_0 $ in the threshold limit $k\to
0$. The energy-independent reduced width $\Gamma_0 = \mu_{\rm res}
\Delta$, where $\mu_{\rm res}$ is the difference between the magnetic
moments of the resonance state and the asymptotically free atoms and
$\Delta$ is the magnetic width of the resonance. The resonance energy is
$E_{\rm res} = \mu_{\rm res} (B-B_0)$, where $B$ is the magnetic
field and $B_0$ is the resonant field. The scattering amplitude
$f(E)$ satisfies the optical theorem \cite{taylor_scattering_1983}.

\begin{figure}[]
	\begin{center}
		\includegraphics[width=0.45\textwidth]{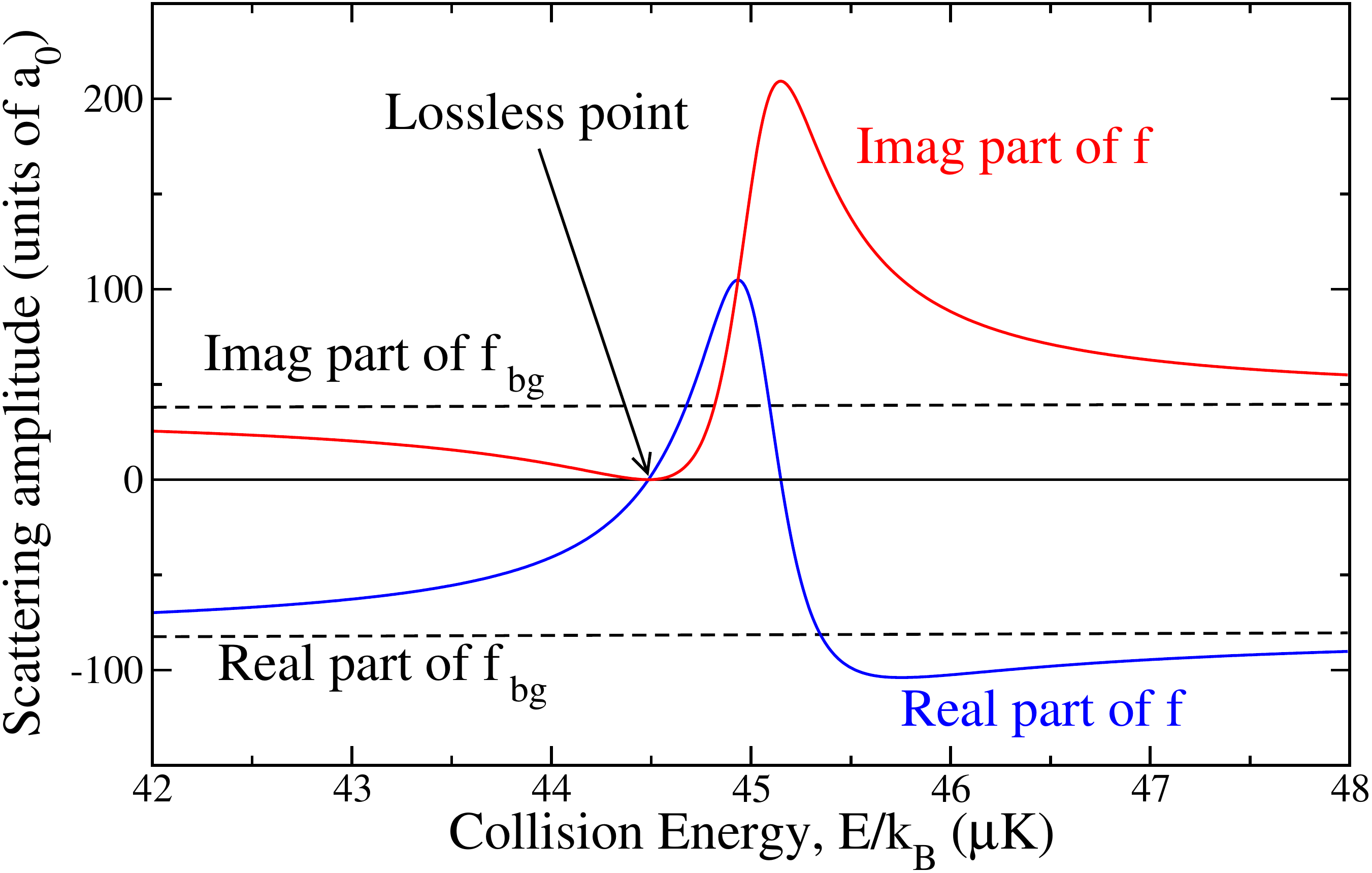}
	\end{center}
	\caption{(Color online) Real and imaginary part of scattering amplitudes, $f$
	and $f_{\rm bg}$, as a function of collision energy for a narrow
	${}^{87}$Rb Feshbach resonance at $B_0= 68.54$ mT and $\Delta
	=0.6$ $\mu$T, with $B-B_0$ tuned at $0.05$ mT. Also shown is the
	lossless point where $\operatorname{Im}f(E)=0$. Values obtained from \cite{chin_feshbach_2010},
	$1 a_0 = 0.0529$ nm, and $k_{\rm B}$ is the Boltzmann constant.}
	\label{fig_sc_amp}
\end{figure}

Figure~\ref{fig_sc_amp} shows $f(E)$
near a Feshbach resonance as a function of collision energy $E$.
The resonance occurs at a finite collision energy and $f(E)$ approaches
$f_{\rm bg}(E)$ away from $E_{\rm res}$. The imaginary part
of $f(E)$ is related to the total cross section $\sigma(E)$ and thus to
the fraction of scattered atoms.  In fact, $\sigma(E)= 4\pi \Im
f(E)/k$ from the optical theorem \cite{taylor_scattering_1983}. 
On resonance Im($f$) is maximal and $\approx 1/k$. More importantly, 
there exists a collision energy at which $f(E)=0$ due to an
interference between the background and resonance scattering amplitudes.
This collision energy, which to good approximation is $E_{\rm res}-\hbar\Gamma_0$ 
for positive $a_{\rm bg}$, is  the lossless optimal
point mentioned in Fig.~\ref{fig_schematic}. 

We now describe the many-body physics of colliding BECs, assuming
that both BECs contain the same atomic species. Then the collision
energy $E \approx E_0/2$.  Their dynamics is well described by the
time evolution of the order parameter $\Psi(\mf{x},t)$, the
expectation value of the annihilation operator $\hat{\Psi}(\mf{x},t)$
in the Heisenberg picture. For
a BEC at rest its evolution is well described by the Gross-Pitaevskii
(GP) equation, derived for an energy-independent and real scattering
amplitude. Both assumptions are invalid near a Feshbach resonance at
finite energy.

Our starting point is Eq.~38 of Ref.~\cite{kohler_microscopic_2002}
obtained using a cumulant expansion. 
It includes the time and energy dependence of the
two-body scattering and is given by
\begin{eqnarray}
  i \hbar \pd{}{t}\Psi(\mf{x},t) &=& H_{\rm 1B}\Psi(\mf{x},t)+ \int
  \prod_i d\mf{y}_i \int_{t_0}^{\infty}dt_1\nonumber \\ && \quad \times
  \Psi(\mf{y}_1,t_1) \Psi(\mf{y}_2,t_1)\Psi^*(\mf{y}_3,t)\label{MB1}\\
  &&\quad\quad\quad\quad\bra{\mf{x},\mf{y}_3}T_{\rm 2B}\left(
  t,t_1\right)\ket{\mf{y}_1 ,\mf{y}_2}\, ,  \nonumber
\end{eqnarray}
where $t_0$ is the initial time, $H_{\rm 1B}=
-\hbar^2\nabla^2/(2m) + V(\mf{x})$, the single-particle Hamiltonian,
and $V(\mf{x})$ is the external potential. The operator $T_{\rm 2B}$
is the two-body $T$-matrix in the time domain and the integrals over
$\mf{y}_i$ for $i =1,2,3$ are in coordinate space.

The momenta of the medium and the laser BEC are
non-overlapping.  Hence, the wave function $\Psi(\mf{x},t)$ is the sum of 
orthogonal medium and laser wavefunctions, $\Psi_{\rm M}(\mf{x},t)$
and $\Psi_{\rm L}(\mf{x},t)$, respectively. Since the laser BEC 
is small in size and number density, we linearize Eq.~\eqref{MB1} for the laser condensate assuming phase-matching conditions. After rearranging 
terms, we find 
\begin{eqnarray}
\label{MB2}
i \hbar \pd{}{t}\Psi_{\rm L}(\mf{x},t) &=& H_{1B}\Psi_{\rm L}(\mf{x},t)+2 \int \prod_i d\mf{y}_i\Psi_{\rm M}^*(\mf{y}_3,t) \nonumber\\ 
&&\times  \int_{t_0}^{\infty}dt_1 \bra{\mf{x},\mf{y}_3}T_{\rm 2B}\left( t,t_1\right)\ket{\mf{y}_1 ,\mf{y}_2}\nonumber \\ 
&&\quad\quad\quad\quad\times \Psi_{\rm L}(\mf{y}_1,t_1) \Psi_{\rm M}(\mf{y}_2,t_1)\, ,
\end{eqnarray}
which is nonlocal in both space and time and ignores the effect
of the laser on the evolution of the medium. Hence, the medium BEC is
described by the ``energy-independent'' GP equation.

A more insightful expression can be obtained when we approximate
the integrands in Eq.~\eqref{MB2} by power series in 
derivatives evaluated at $\bf x$ and $t$.  First, we
realize that $T_{\rm 2B}(t,t_1)$ only depends on $t-t_1$ and is peaked
around $t-t_1 =0$. Assuming that the wave functions vary slowly in time, the lower limit
of the integral over time can be extended to $-\infty$. Next, we note
for any pair of functions $g(t)$ and $h(t)$
\begin{equation}
	\int_{-\infty}^{\infty} d\tau h(\tau)g(t-\tau) = \tilde{h}\left( i\p/\p t \right)g(t)\, ,
	\label{eq_trick}
\end{equation}
where $\tilde{h}(z)= \int dt\, e^{iz t} h(t)$ is the Fourier transform of $h(t)$ and 
$\tilde{h}(i\p/\p t)=\sum_n d^n\tilde{h}/dz^n\mid_{z=0}(i\p/\p t)^n/n!$.
Using Eq.~\eqref{eq_trick} with $h(\tau)=\bra{.}T_{\rm 2B}\left( \tau\right)\ket{.}$, the interaction term in Eq.~\eqref{MB2} reduces to 
\begin{align}
  &  2 \int \prod_i d\mf{y}_i  \Psi_{\rm M}^*(\mf{y}_3,t) 
  \label{eq_int}\\
  &\quad\times\bra{\mf{x},\mf{y}_3}T_{\rm 2B}\left(i\hbar\p/\p t\right)\ket{\mf{y}_1 ,\mf{y}_2}\Psi_{\rm L}(\mf{y}_1,t) \Psi_{\rm M}(\mf{y}_2,t)\, , \nonumber
\end{align}
where the $T$-matrix $T_{\rm 2B}(z)$ is now in the energy domain (dropping
the $\sim$ notation for simplicity) and the time derivatives only act
on $ \Psi_{\rm L}(\mf{y}_1,t) \Psi_{\rm M}(\mf{y}_2,t)$.

The $T$-matrix in coordinate space can be evaluated
by transforming to the momentum representation. 
For $s$-wave scattering the dependence on the relative
momenta can be neglected. That is, to a good approximation the $T$-matrix
in momentum representation is \cite{kohler_microscopic_2002}
\begin{eqnarray}
	\lefteqn{ \bra{\mf{k}_4,\mf{k}_3}T_{\rm 2B}\left(z\right)\ket{\mf{k}_1 ,\mf{k}_2}
	= \delta(\mf{k}_4+\mf{k}_3-\mf{k}_2-\mf{k}_1)}\nonumber\\
  &&\times -\frac{\hbar^2}{4\pi^2 \mu} f\left(z-\hbar^2(\mf{k}_1+\mf{k}_2)^2/2M\right)\, ,
  \label{eq_t_mat}
\end{eqnarray}
where the $\delta$-function reflects total momentum conservation, $f(E)$ is
the scattering amplitude of Eq.~\eqref{eq_sc_amp}, and $M =2m$.

Inserting the momentum representation of $T_{\rm 2B}$
into Eq.~\eqref{eq_int} and noting formally that $\phi(\mf{y},t)=
\exp[-i(\mf{y}-\mf{x})\cdot i\nabla]\phi(\mf{x},t)$, the Taylor
expansion of $\phi(\mf{y},t)$ around position $\mf{x}$, the 
interaction term becomes
\begin{align}
	&\frac{2}{(2\pi)^6}\left( -\frac{\hbar^2}{4\pi^2 \mu} \right)\iint 
	\prod_{i,j} d\mf{y}_i d\mf{k}_j \delta(\mf{k}_4+\mf{k}_3-\mf{k}_2-\mf{k}_1)\nonumber\\
&\quad\times	\Psi_{\rm M}^*(\mf{y}_3,t)e^{iQ}
       f\left( i\hbar\p/\p t- \hbar^2(\mf{k}_1+\mf{k}_2)^2/2M \right) \nonumber\\
   &\quad \times    \Bigl[ e^{-i \mf{(y_1-x)}\cdot i\nabla} \Psi_{\rm L}(\mf{x},t)\Bigr]
  \Bigl[ e^{-i \mf{(y_2-x)}\cdot i\nabla} \Psi_{\rm M}(\mf{x},t)\Bigr],
	\label{}
\end{align}
where $Q=\mf{k}_4\cdot \mf{x}+\mf{k}_3\cdot \mf{y}_3-\mf{k}_2\cdot \mf{y}_2-\mf{k}_1\cdot \mf{y}_1$. 
Performing all integrations,  
we find a local equation of motion for $\Psi_{\rm L}(\mf{x},t)$. That is,
\begin{eqnarray}
\lefteqn{ i\hbar \pd{}{t}\Psi_{\rm L}(\mf{x},t) = H_{1B}\Psi_{\rm L}(\mf{x},t)
-\frac{4\pi\hbar^2}{\mu} \Psi_{\rm M}^*(\mf{x},t)}
	\label{eq_int_final}\\
  &&\times f\left( i\hbar\p/\p t +\hbar^2\nabla^2/2M \right)\Psi_{\rm L}(\mf{x},t)\Psi_{\rm M}(\mf{x},t)\, .
  \nonumber
\end{eqnarray} 

\begin{figure}
	\begin{center}
		\includegraphics[width=0.45\textwidth]{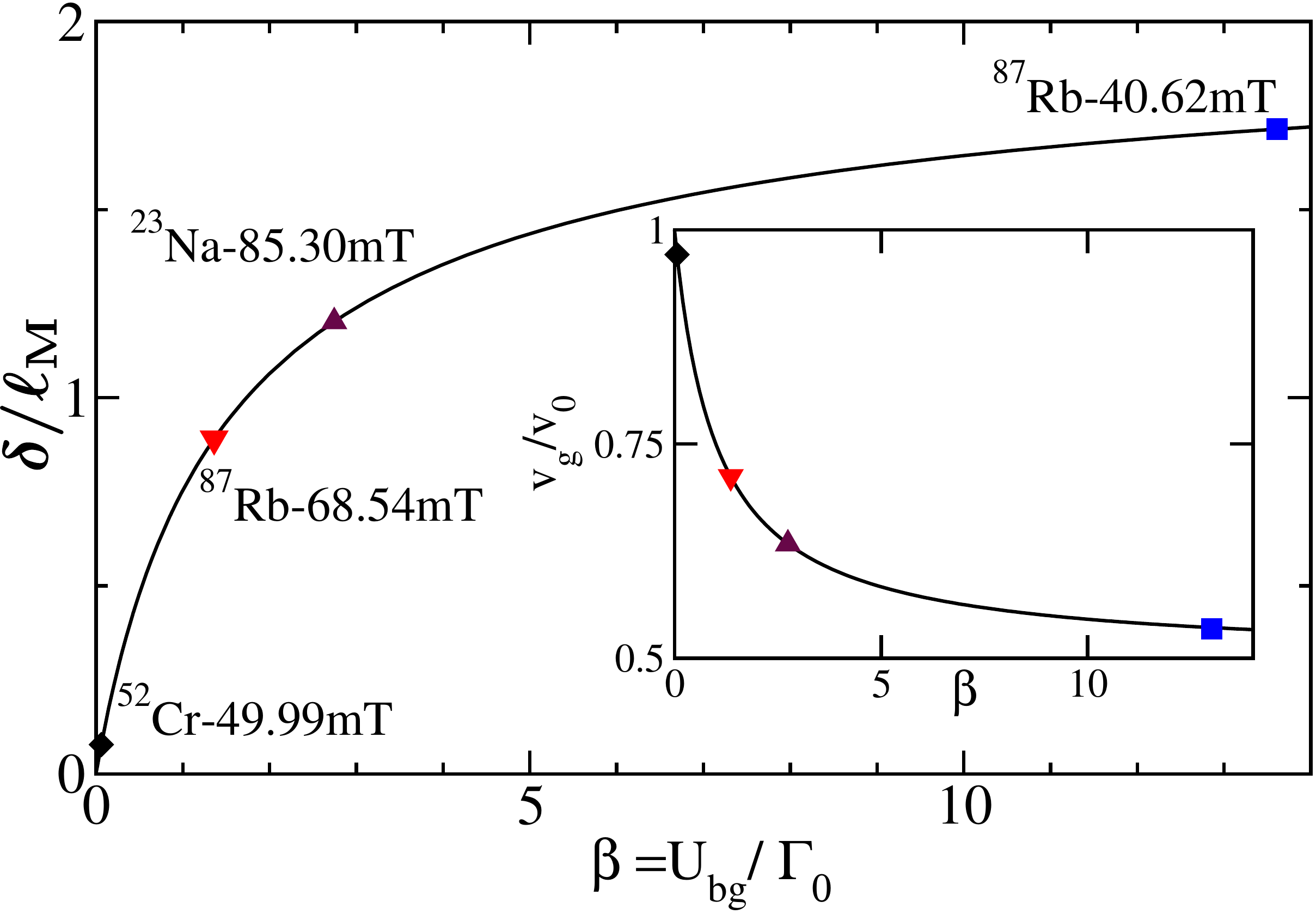}
	\end{center}
	\caption{(Color online) Distance delay $\delta$ of the laser
	BEC normalized by the Thomas-Fermi radius $\ell_{\rm M}$ of
	the medium BEC as a function of dimensionless parameter $\beta
	= U_{\rm bg}/\Gamma_0$,  where $U_{\rm bg}$ and $\Gamma_0$
	are defined in the text.  The delay for selected resonances
	assuming a peak number density of the medium of $n_{\rm M} =
	10^{15}\,{\rm cm}^{-3}$ is shown by colored markers.
	The inset shows the group velocity $v_{\rm g}$ of the laser
	 BEC in a homogeneous medium BEC
	as a function of $\beta$.  Here $v_0$ is the free space velocity of the
	laser BEC.  Markers indicate $v_{\rm g}$ for the
	same selected resonances and $n_{\rm M}$ as in the main figure.
	 }
	\label{fig_delta}
\end{figure}

Since $\ell_{\rm L}\ll \ell_{\rm M}$ and the spread in the collision
energy is much smaller than $\Gamma(E_0/2)$, it is sufficient to expand
$f(z)$ to first order around $z=E_0/2$, the average relative collision
energy, and derivatives of $\Psi_{\rm M}(\mf{x},t)$ can
be neglected.  For our homonuclear system the time evolution of the
laser condensate is then given by
\begin{eqnarray} 
	i\hbar \pd{}{t}\Psi_{\rm L}(\mf{x},t)  &=&
	\Bigl[-\frac{\hbar^2}{2m^*(\mf{x})}\nabla^2+V_{\rm mf}(\mf{x})\nonumber\\
	&&\quad\quad+\,V_{\rm deriv}(\mf{x})\Bigr]\Psi_{\rm L}(\mf{x},t) \, , \label{eq_MB_final}
\end{eqnarray}
where $m^*(\mf{x})= m[1+2\alpha(\mf{x})]/[1+\alpha(\mf{x})]$ is the
position-dependent effective mass and $\alpha(\mf{x}) =(4\pi \hbar^2/m)
|\Psi_{\rm M}(\mf{x})|^2 df(z)/dz$ with $df(z)/dz$ evaluated at $z=
E_0/2$. The ``mean-field'' potential $V_{\rm mf}(\mf{x})= [V({\bf x}) -
(8\pi \hbar^2/m)|\Psi_{\rm M}(\mf{x})|^2f(E_0/2)]/[1+2\alpha(\mf{x})]$
contains the external potential and a potential proportional to the 
scattering amplitude and medium density.  The latter contribution
 is analogous to the interaction potential in
the GP equation except that the scattering amplitude is evaluated
at non-zero energy $E_0/2$. Finally, the potential $V_{\rm
deriv}(\mf{x})=E_0\alpha(\mf{x})/(1+2\alpha(\mf{x}))$.  The factor
$1+2\alpha(\mf{x})$, appearing throughout, results from the $i\hbar
\p/\p t$ argument of the scattering amplitude.

The operator acting on $\Psi_{\rm L}(\mf{x})$ on the right-hand side
of Eq.~\eqref{eq_MB_final} is not Hermitian as the scattering amplitude
is complex valued.  In fact, the non-Hermicity leads to atom loss out
of both condensates, shown in Fig.~\ref{fig_schematic}b as the halo.
 For a medium number
density $n_{\rm M}$,
the loss rate out of the laser condensate is $n_{\rm
M}v_0 \sigma$.  Consequently, at resonance, where 
 $\sigma \approx 8\pi/(k_0/2)^2$, 
the fraction of atoms remaining in the laser condensate after the collision
is $\approx \exp\left(-8\pi n_{\rm
M} \ell_{\rm M}/(k_0/2)^2 \right)$. For typical values of $n_{\rm M}$
and $\ell_{\rm M}$ almost all the laser atoms are lost at resonance.

For our proposal we need to minimize these losses.
We can use the lossless point where $f(z)=0$, indicated in
Fig.~\ref{fig_sc_amp}, and the total cross section is zero.  The effective
mass and the potentials in Eq.~\eqref{eq_MB_final} are
then real.  In fact, $m^*(\mf{x})>m$ and $df/dz = a_{\rm bg}/\Gamma_0$.

At the lossless point the simplest case to analyse is that of a homogeneous medium and $V({\bf x})=0$. 
The potential $V_{\rm mf}(\mf{x})$ vanishes and the effective mass
is uniform.  Transforming Eq.~\eqref{eq_MB_final} to momentum space,
we find that the propagation or group velocity of the laser BEC is
\begin{equation}
	\mf{v}_{\rm g}(\mf{k}_0)\Big|_{\rm loss\, less}=\frac{\hbar \mf{k}_0}{m^*} =
	{\bf v}_0 \left[ \frac{1+\beta/2  \, } {1+\beta}\right]\, ,
  \label{eq_group}
\end{equation} 
where the dimensionless quantity $\beta = U_{\rm bg}/\Gamma_0>0$ and
$U_{\rm bg}= (8\pi \hbar^2/m) a_{\rm bg}n_{\rm M}$ is the background
mean-field interaction energy. The inset of Fig.~\ref{fig_delta} shows
the group velocity as a function of $\beta$. The group velocity lies
between ${\bf v}_0/2$ and ${\bf v}_0$ and approaches ${\bf v}_0/2$
when $\beta\to\infty$.

We now turn to propagation through an inhomogeneous medium, but still with
$V({\bf x})=0$. The assumption $\ell_{\rm L}\ll \ell_{\rm M}$ implies that
the density variation of the medium orthogonal to the laser propagation
direction is negligible and we only need to treat propagation along
$\mf{k}_0$ passing through the center of the medium BEC.

For simplicity the density profile of the untrapped medium is given
by $|\Psi_{\rm M}(\mf{x})|^2 = n_{\rm M}(1-x^2/\ell_{\rm M}^2)$,
using the Thomas-Fermi approximation and assume that expansion of the
medium can be neglected. Here $n_{\rm M}$ is the peak number density and
$\ell_{\rm M}$ is the Thomas-Fermi radius of the medium. We assume
that $V_{\rm deriv}(\mf{x})\ll E_0/2$ for all $\mf{x}$ and, hence, can
apply the Wentzel-Kramers-Brillouin (WKB) approximation to estimate
$\delta$. We find
\begin{equation}
	\delta/\ell_{\rm M} = 2\left(1 - 
	\frac{\operatorname{arctanh}\left( \sqrt{\beta/(1+\beta)} \right)}{\sqrt{\beta(1+\beta)}} \right)\,,
	\label{eq_delta}
\end{equation}
and the dimensionless quantity $\beta = U_{\rm bg}/\Gamma_0$ is evaluated
at the peak number density $n_{\rm M}$. Figure \ref{fig_delta} shows
$\delta$ as a function of $\beta$.  The maximum $\delta$ that can be
attained by the laser is $2\ell_{\rm M}$ for $\beta \to \infty$.

There are several constraints on the realization of the proposal. Firstly, we have $\ell_{\rm
L}\ll \ell_{\rm M}$. Secondly, scattering is  $s$-wave dominated, so that 
$k_0 a_{\rm bg}\ll 1$ or $E_0/2 \ll \hbar^2/(2ma_{\rm bg}^2)\equiv
E_{\rm bg}$, the Wigner threshold limit. 
Thirdly, by solving for $\Im f(z)=0$ for positive $a_{\rm bg}$,
we find that the requirement $\Gamma_0< E_{\rm res}$ must hold.  Fourthly, the energy window
around the lossless point, where $\Im f$ is small, is on the order of
$\Gamma(E_{\rm res})$. Consequently, the energy width of the laser BEC,
$\Delta E_L \approx \hbar^2k_0/(2m\ell_{\rm L})$ 
, must satisfy $\Delta E_L \ll \Gamma(E_{\rm res})$.  In other
words  $\ell_{\rm L} \gg \hbar^2/(m a_{\rm bg}\Gamma_0)\equiv \ell_{\rm
L}^{\rm min}$. Finally, we require resonances 
for which $\delta$ is comparable or larger than the size of the laser BEC. 
Since $\ell_{\rm L} \ll \ell_{\rm M}$, we have $\beta=U_{\rm bg}/\Gamma_0$ 
is at least of order one.

Table \ref{tab_shift} gives a non-exhaustive list of narrow resonances, 
which satisfy the constraints. For four of these resonances
the expected $\delta$ is shown in Fig.~\ref{fig_delta} assuming a peak
density of $n_{\rm M}=10^{15}$ cm$^{-3}$.  If we assume $\ell_{\rm
L}/\ell_{\rm M} \approx 0.1$ then $\delta$ ranges from  $0.1 \ell_{\rm
L}$ to $20 \ell_{\rm L}$ for the resonances in Table \ref{tab_shift}. For
the selected density the chromium resonance is only a marginal candidate
for slowing experiments. 

\begin{table}[]
	\caption{
Resonance parameters, experimental constraints, and spatial delay for
nine Feshbach resonances. The first five columns specify the Feshbach
resonance.  The columns are the atomic species, magnetic resonance
position $B_0$, magnetic width $\Delta$, reduced width $\Gamma_0$, and
Wigner-threshold limit $E_{\rm bg}$. 
The sixth column gives the minimum size, $\ell_{\rm L}^{\rm min}$,
of the laser BEC.  The last column is the shift $\delta$ in units of
the radius of the medium $\ell_{\rm M}$, assuming a peak medium density
of $n_{\rm M}=10^{15}$ cm$^{-3}$. Parameters obtained from
\cite{chin_feshbach_2010}.  
}
	\begin{tabular}{c c cccccccc}
	\hline
	\hline
	Atom  & $B_0$ & $\Delta$ &  
	$\Gamma_0/k_{\rm B}$ & $E_{\rm bg}/k_{\rm B}$ &$\ell_{\rm L}^{\rm min}$ & $\delta/\ell_{\rm M}$\\
		& (mT) 	& (mT)	& ($\mu K$) & ($\mu K$) & ($\mu m$) & \\
	\hline
	$^{23}$Na  & 119.5 & $-0.14$  &  14 &1900 & 0.45 & 0.15\\
	$^{\pr\pr}$  & 90.7 & 0.10  &260& 1900 & 0.025 & 0.0091\\
	$^{\pr\pr}$  & 85.3 &  $2.5\times 10^{-4}$ & 0.64 & 1900 & 9.8 & 1.20\\
	$^{87}$Rb  & 100.74 & 0.021 & 39& 200& 0.027& 0.025\\
	$^{\pr\pr}$& 91.17 &  $1.3 \times 10^{-4}$ &  0.24 & 200 & 4.4 & 1.25\\
	$^{\pr\pr}$&  68.54 & $6 \times 10^{-4}$ & 0.54 &200  & 1.9 & 0.89\\
	$^{\pr\pr}$&  40.62 & $4 \times 10^{-5}$ & 0.054 & 200  & 19 &1.7\\
	$^{\pr\pr}$&  9.13 & $1.5 \times 10^{-3}$&  2.0 & 200 & 0.52 & 0.38\\
	$^{52}$Cr &  49.99 & 0.008 &  22 & 290 & 0.076 &0.078 \\
	\hline
	\hline
\end{tabular}
\label{tab_shift}
\end{table}

In conclusion we have shown that collisions in the presence of a
magnetic Feshbach resonance can lead to slowing of a laser BEC as it
propagates through a large medium BEC.  The slowing is a consequence of
the collision-energy dependence of the scattering amplitude near the
resonance.  Based on a generalized Gross-Pitaevskii equation,
we predict a maximal reduction of the group velocity by a factor of two
and suggest that the experiment be performed at a magnetic field where
the elastic scattering  is zero. Such a field always
exists near a magnetic Feshbach resonance. For finite-sized condensates
slowing can be observed by measuring the spatial 
delay of the laser BEC, which can not exceed twice the Thomas-Fermi
radius of the medium. We show that for narrow resonances this
signal is expected to be measurable.

We acknowledge support from a Physics Frontier Center of the National Science Foundation located at the Joint Quantum Institute. We acknowledge valuable discussions with Noah Bray-Ali.
\bibliography{ref_slow_atoms}

\end{document}